\documentclass[manuscript]{acmart}

\AtBeginDocument{%
  \providecommand\BibTeX{{%
    \normalfont B\kern-0.5em{\scshape i\kern-0.25em b}\kern-0.8em\TeX}}}

\copyrightyear{2024}
\acmYear{2024}
\setcopyright{acmlicensed}\acmConference[XAIxArts 2024]{Explainable AI for the Arts Workshop 2024}{June 23, 2024}{Chicago, IL, United States}
\acmBooktitle{2nd International Workshop on Explainable AI for the Arts (XAIxArts), ACM Creativity and Cognition (C\&C '24), June 23, 2024, Chicago, USA}

\begin{document}

\title{Explainability Paths for Sustained Artistic Practice with AI}

\author{Austin Tecks}
\email{austin.tecksbleuer@concordia.ca}

\author{Thomas Peschlow}
\email{thomas.peschlow@concordia.ca}

\author{Gabriel Vigliensoni}
\affiliation{
  \institution{Concordia University}
  \streetaddress{1455 De Maisonneuve Blvd. West}
  \city{Montréal}
  \country{Canada}
  \postcode{H3G 1M8}}
\email{gabriel.vigliensoni@concordia.ca}

\renewcommand{\shortauthors}{Tecks, Peschlow, and Vigliensoni}

\begin{abstract}
The development of AI-driven generative audio mirrors broader AI trends, often prioritizing immediate accessibility at the expense of explainability. Consequently, integrating such tools into sustained artistic practice remains a significant challenge. 
In this paper, we explore several paths 
to improve explainability, drawing primarily from our research-creation practice in training and implementing generative audio models. As practical provisions for improved explainability, we highlight human agency over training materials, the viability of small-scale datasets, the facilitation of the iterative creative process, and the integration of interactive machine learning as a mapping tool.
Importantly, these steps aim to enhance human agency over generative AI systems not only during model inference, but also when curating and preprocessing training data as well as during the training phase of models.

\end{abstract}



\maketitle

\section{Introduction}
The rapid proliferation of pre-trained generative audio models belies the minimal adoption these tools have seen as part of sustained sound and music practices. In step with prevailing AI trends, popular tools like Suno\footnote{\href{https://suno.com/}{\texttt{https://suno.com/}}} and Stable Audio\footnote{\href{https://stableaudio.com/}{\texttt{https://stableaudio.com/}}} have adopted text-based user interfaces favouring broad appeal and immediate access at the expense of human agency and interpretability \cite{lauriolaIntroductionDeepLearning2022}. 
This trend is highlighted by the widespread use of natural language conditioning, which, despite broadening user access, presents limitations for sound artists and musicians who require precise control to maintain long-term temporal coherence in the sonic output of generative systems \cite{vigliensoni23steering}. 
Moreover, in a bid to offer wide-ranging user experiences, generative audio models such as Jukebox and Suno have been trained on unfathomably large datasets, the former boasting a dataset of more than 1.2 million songs \cite{dhariwalJukeboxGenerativeModel2020}. Although this may facilitate the generation of novel and unexpected hybrids across music genres and provide satisfying responses to common prompts, the practical value of these innovations for artists seeking a sustained practice remains questionable.

While the conditions presented above have successfully piqued the public's interest in generative AI art, our ongoing research-creation practice with neural audio synthesis illuminates alternative paths, prioritizing explainability for sustained artistic practice. This paper highlights our work preparing intimately curated, small-scale datasets for training and implementing neural audio synthesis models with RAVE \cite{caillonRAVEVariationalAutoencoder2021}. Throughout this process of data curation and preprocessing, training, and model inference, we have identified the following provisions for improved explainability:

\begin{itemize}
\item Improving agency through human-scale models and artist-curated datasets
\item Extending the iterative process beyond inference to curation and training
\item Defining the performance space through interactive machine learning  \cite{failsInteractiveMachineLearning2003}
\end{itemize}

In Sections \ref{sec:human}, \ref{sec:iteration}, and \ref{sec:IML}, we will define the above-mentioned provisions, and in Section \ref{sec:case}, we will draw from our research-creation practice, illustrating the potential of explainable AI (XAI) to reconcile the opaqueness of generative AI with the demands of sustained artistic practice.

\section{Human-scale models, Artist-curated Datasets}\label{sec:human}
The suitability of AI-driven generative audio for sustained creative work depends on artist control over training material, which in turn often relies on the viability of smaller-scale datasets. While mainstream machine learning models, pre-trained for public use, thrive on large amounts of data, artists usually find greater utility in models with a more narrow scope, facilitating a deeper connection to the training material as well as the ability to more effectively steer the model towards a creative objective \cite{vigliensoni22small}. 
Notable examples of extremely narrow and focused datasets are Holly+, a sonic digital likeness of Holly Herndon capable of reconstructing her voice \cite{freethinkAIChangingMusic2023}, and the early work by Dadabots, where single albums are used as training data to generate music “within the limited aesthetic space of the album’s sound”~\cite{carr18generating}.
Similarly, using smaller datasets can enhance transparency, agency, and bolster an artist's confidence and trust in a model, thereby addressing a major challenge artists and the public face in adopting AI-driven technologies~\cite{choung2023trust}.

\section{Extending the iterative process beyond inference}
\label{sec:iteration}

Iteration is an essential process for artistic development. It facilitates the emergence of novel and meaningful insights \cite{chanImportanceIterationCreative2015} and is a powerful force for creativity \cite{sawyerIterativeImprovisationalNature2021}.
Perhaps the most appealing affordance provided by pre-trained models is the ability to jump straight into an iterative, inferential process within a satisfactory time frame, a process enabled by externalized computational power. This experience, however, usually confines the creative process to iterative conditioning on text prompts. Furthermore, artists may find such a model inadequate in providing the conditions required for fruitful creative iteration due to a lack of fine control of the generative process, its intrinsic non-causality, and the lack of long-term temporal coherence.
We propose that artists would benefit most from machine learning models that support an iterative process both during the inference and training phases. This provision is interdependently related to the scale of the model, as smaller-scale datasets potentially yield shortened training times.

\section{Defining the performance space through Interactive Machine Learning}\label{sec:IML}
We also link explainability in creative AI practices to a practitioner's ability to steer the models during performance. At training time, artists guide the learning process by curating the data for training the system and by continuously observing and adjusting the process. However, at inference time, due to the stochastic nature of training, artists may have to perform with models whose axes and parameters are unknown. To address this issue, we propose a regressive approach where we map the human performance space to the computer's latent space, using interactive machine learning as a mapping tool~\cite{vigliensoni23steering}. This method involves exploring the latent space, identifying points of interest, and mapping these to specific points in the performance space. By doing so, we define the space rather than merely explaining it. This strategy has proven effective for real-time interaction with generative models, even those that are highly dimensional.

\section{Case Study}\label{sec:case}
As a case study, we provide insights derived from our experience training generative audio models on a dataset composed of archival recordings from the Museo de la Memoria y los Derechos Humanos in Santiago, Chile.\footnote{\href{https://archivoradial.museodelamemoria.cl/}{\texttt{https://archivoradial.museodelamemoria.cl/}}}
In our overview, we examine the crucial creative steps in the development of this project: data-preparation, training, and performance.

\subsection{Data Preparation}
Data preparation entails curating, classifying, and normalizing a given dataset. This process initially precedes training but exists within an iterative cycle of data preparation, training, and implementation, thus providing ample opportunity for applied human agency. In preparing our data, we curated three separate datasets from about 35 hours of viable audio recordings. These datasets were organized based on their historical and semantic content as well as their sonic coherence, resulting in categories of public recordings, music, and home recordings. 
From here, the recordings were processed and normalized to enhance fidelity and achieve a baseline of intra-dataset spectral congruency. 

Normalizing the audio data and sequestering the recordings into separate datasets according to their sonic characteristics facilitate an optimal training process. If done effectively, this process can improve the reliability of a generative model, allowing us to better anticipate its behavior, an essential component for the foundation of trust and transparency~\cite{10.1145/3442188.3445923}. Relatively short training cycles enable us to alter the decisions made at this stage with each training iteration, effectively compounding human agency within the data preparation process.

\subsection{Training}\label{sec:training}

RAVE’s training process, composed of clearly segmented phases with distinct task orientations, allows for meaningful alterations to the model over the iterative process, further enabling human agency. The first training phase is based on a variational autoencoder (VAE) \cite{vae}. It focuses on the mathematical optimization of a compression and decompression process in which the size of the bottleneck, or the dimensionality of the latent space, is an important hyperparameter that can both be set by the user, or automatically derived by the model by virtue of the model being what is called a disentangled variational autoencoder \cite{disentangled}. The second training phase is based on a generative adversarial network (GAN)~\cite{gan}. This phase focuses on the perceptual optimization of the output by fine-tuning the model through a process consisting of translating the encoding of noisy data into sound pertaining to the original dataset.

For the VAE phase, a 5 million-step training phase was deemed optimal for all three datasets despite their sonically distinct nature. This may be due to the shared prominence of the human voice and the relative absence of discontinuous transient-rich material.
In the GAN phase, we maintain consistency between datasets, ranging from one to two million steps. This phase often presents the most significant fidelity improvements. However, prolonged training in this phase can produce a smoothing effect on output, potentially diminishing certain sound characteristics and introducing out of domain sonic artifacts to generation output.

\subsection{Performance}\label{sec:performance}

One feature that sets RAVE apart from other generative audio models is its ability to perform model inference in real time. We chose a latent space size of eight dimensions to provide a nuanced understanding of each axes' impact on output while allowing diverse sound reconstruction.
We have experimented with various gestural interfaces to control RAVE models. For this project, we chose to use our face as the performance space, utilizing Google's MediaPipe Face Mesh model\footnote{\href{https://developers.google.com/mediapipe/solutions/vision/face\_landmarker/index}{\texttt{https://developers.google.com/mediapipe/solutions/vision/face\_landmarker/index}}} to embody this interaction. We use the interactive machine learning approach to map our facial landmarks and movements to the model's latent space. A video demonstrating the degree of steerability achieved with the provisions stated in this paper can be watched at \href{https://media.vigliensoni.com/video/xaixarts2024}{\texttt{\smaller{https://media.vigliensoni.com/video/xaixarts2024}}}.

\section{Conclusion}\label{sec:conclusion}
Through our research-creation practice, we have identified reliable methods to enhance explainability and steerability throughout all stages of interaction with a generative audio model. Smaller datasets, supported by models like RAVE, allow artists to work with datasets with which they are intimately familiar. Extending the iterative process to training compounds artist agency and improves a model's explainability. Additionally, the inherent iterative nature of interactive machine learning and its mapping capabilities enable artists to define axes and zones for exploration in generative models. This allows them to create causal gestures and produce sound and music with long-term temporal coherence. We believe that sustained artistic practice benefits significantly from the explainable pathways these provisions supply.

\begin{acks}
This project has been funded by Concordia University's FRDP grant. Important parts of this work used Digital Research Alliance of Canada’s High Performance Computing resources.
\end{acks}

\bibliographystyle{ACM-Reference-Format}
\bibliography{references}

\end{document}